\RequirePackage{fix-cm}
\documentclass[smallcondensed]{svjour3}     
\usepackage[dvipdfmx]{graphicx}
\usepackage{amsmath}
 \journalname{eps}
\begin{document}

\title{Interplanetary particle transport simulation for warning system for aviation exposure to solar energetic particles}

\titlerunning{Interplanetary particle transport simulation}        

\author{Y\^uki Kubo         \and
         Ryuho Kataoka \and Tatsuhiko Sato
}


\institute{Y\^uki Kubo \at
              National Institute of Information and Communications Technology, Tokyo 184-8795, Japan \\
              \email{kubo@nict.go.jp} 
           \and
           Ryuho Kataoka \at
              National Institute of Polar Research, Tokyo 190-8518, Japan \\
Department of Polar Science, The Graduate University for Advanced Studies (SOKENDAI), Tokyo 190-8518, Japan \\
	\and
	Tatsuhiko Sato \at
	Japan Atomic Energy Agency, Ibaraki 319-1195, Japan
}

\date{Received: date / Accepted: date}

\maketitle

\begin{abstract}
Solar energetic particles (SEPs) are one of the extreme space weather phenomena. A huge SEP event increases the radiation dose received by aircrews, who should be warned of such events as early as possible. We developed a warning system for aviation exposure to SEPs. This article describes one component of the system, which calculates the temporal evolution of the SEP intensity and the spectrum immediately outside the terrestrial magnetosphere. To achieve this, we performed numerical simulations of SEP transport in interplanetary space, in which interplanetary SEP transport is described by the focused transport equation. We developed a new simulation code to solve the equation using a set of stochastic differential equations. In the code, the focused transport equation is expressed in a magnetic field line coordinate system, which is a non-orthogonal curvilinear coordinate system. An inverse Gaussian distribution is employed as the injection profile of SEPs at an inner boundary located near the Sun. We applied the simulation to observed SEP events as a validation test. The results show that our simulation can closely reproduce observational data for the temporal evolution of particle intensity. By employing the code, we developed the WArning System for AVIation Exposure to Solar energetic particles (WASAVIES).

\end{abstract}

\section{Introduction}
\label{sec:introduction}
Space radiation poses a serious threat to several human activities, such as high radiation doses for astronauts, adverse effects on aircrew health, artificial satellite malfunctions, and the disruption of high-frequency radio wave communications in high-latitude regions. Predicting the space radiation environment to reduce the risk of radiation hazards is one of the most important goals of space weather research. \par

Space radiation is primarily composed of energetic protons and electrons. Radiation doses from the protons are harmful for astronauts and aircrews, whereas the electrons predominantly affect artificial satellites. Energetic protons originate from galactic cosmic rays (GCRs) and solar energetic particles (SEPs). The spectrum and intensity of GCRs, and hence their radiation doses, are almost constant over time scales of much less than the 11 year solar cycle. On the other hand, SEPs occur sporadically and their intensity suddenly increases by multiple orders of magnitude within days. Such sudden increases in SEP intensity present a major radiation hazard to astronauts, adversely affecting the success of space missions [{\it Hu et al.}, 2009]. While low-energy SEPs cannot penetrate a spacecraft or extravehicular activity suits, high-energy SEPs, the energies of which exceed several tens of MeV, penetrate these materials [{\it Kronenberg and Cucinotta}, 2012; {\it Reames}, 2013]. Therefore, predicting high-energy SEP events is the first priority in ensuring the safety of astronauts and the success of space missions. For an excellent review on the space radiation environment, refer to Vainio {\it et al.} [2009]. \par

Despite the lack of understanding of the SEP mechanism, many researchers have attempted to predict the occurrence of SEP. However, a physics-based definitive prediction remains difficult because SEP occurrence is related to various physical processes yet to be fully clarified, such as solar flares, coronal mass ejections, coronal and interplanetary shock waves, and particle acceleration and transport mechanisms. Therefore, previous studies on predicting the occurrence of SEP have adopted statistical, empirical, or probabilistic approaches [e.g., {\it Garcia}, 2004a, 2004b; {\it Kubo and Akioka}, 2004; {\it Kuwabara et al.}, 2006; {\it Kahler et al.}, 2007; {\it Posner}, 2007; {\it Balch}, 2008; {\it Laurenza et al.}, 2009; {\it N\'u\~nez}, 2011]. \par

On the other hand, {\it Reames} [2004] stated that, ``In the author's opinion, reliable predictions of the onset and fluence of an SEP event prior to its occurrence are not likely in our lifetime. However, after an event onset, it should be possible to model the CME, the shock, and the acceleration and transport of particles sufficiently well to predict the peak intensity at shock passage and the event fluence." Some researchers have focused on simulating the temporal evolution of SEP intensity, and numerous simulation studies of particle acceleration and transport in interplanetary space have been published [e.g., {\it Lario et al.}, 1998; {\it Zank et al.}, 2000; {\it Rice et al.}, 2003; {\it Sokolov et al.}, 2004; {\it Verkhoglyadova et al.}, 2009]. However, the aim of most of these studies was not to predict the SEP intensity profile. To the best of our knowledge, the only operational prediction model that incorporates the physical mechanism of SEPs was developed by {\it Aran et al.} [2006], who predicted the SEP intensity profile from numerical simulations of interplanetary shock propagation and energetic particle transport. Their model is specific to low-energy SEP intensity profiles, the energy of which is less than several tens of MeV. \par

A ground level enhancement (GLE), which is one of the extreme space weather phenomena, is induced by extremely energetic SEPs having energies greater than 450 MeV (approximately 1 GV in rigidity for protons) [{\it Shea and Smart}, 2012]. Energetic SEPs can produce secondary particles such as neutrons, muons, neutrinos, and gamma rays through nuclear reactions in the terrestrial atmosphere. All such secondary particles induce a marked increase in the radiation dose rate at aviation altitudes. This means that effort should be made to alert aircrews to the presence of extremely energetic SEPs so that they can take necessary action to reduce the radiation dose rate they are exposed to. Recently, {\it Kataoka et al.} [2014] developed the WArning System for AVIation Exposure to Solar energetic particles (WASAVIES). The system aims to warn aircrews of high dose levels and to simulate GLEs by forward modeling from SEP transport to an air shower in the terrestrial atmosphere [{\it Sato et al.}, 2014]. \par

This article describes a newly developed numerical simulation code to calculate SEP transport in interplanetary space for the WASAVIES system. The next section describes the SEP transport equation and a coordinate system to express the equation, along with its solution method. Then, the results of test calculations are described. Subsequently, comparisons of calculations with observational data are presented, which are followed by a discussion section. The final section presents a summary. \par

\section{Numerical simulation of solar energetic particle transport}
\label{sec:numerical}
SEPs are believed to be accelerated within three regions: the reconnection region in solar flares, coronal shock waves, and interplanetary shock waves. 
According to {\it Caprioli and Spitkovsky} [2014], the exponential cutoff energy of an ion energy spectrum produced by diffusive shock acceleration, which roughly represents the maximum energy achievable, is proportional to the background magnetic field intensity. Thus, the majority of energetic SEPs cannot be accelerated by interplanetary shock waves, for which the background magnetic field is weak. This means that they tend to be accelerated near the Sun (by solar flares and/or coronal shocks) and then transported to the Earth along interplanetary magnetic fields. Therefore, energetic SEP events will be reproduced by solving the SEP transport equation from the Sun to the Earth if an accelerated energetic SEP time profile near the Sun is assumed. In this section, we describe the numerical simulation of SEP transport through interplanetary space.

\subsection{Focused transport equation}
\label{sec:transport_eq}
As the scattering mean free path of interplanetary SEP transport is roughly comparable to the distance between the Sun and the Earth, SEPs are not isotropically distributed in interplanetary space, particularly in the initial phase of SEP events. Therefore, interplanetary SEP transport cannot be modeled using Parker's transport equation, which assumes an isotropic particle distribution. Instead, the SEP transport equation must explicitly incorporate changes in pitch-angle. For this reason, a focused transport equation is widely used to simulate interplanetary SEP transport [e.g., {\it Ruffolo}, 1995; {\it Zhang et al.}, 2009; {\it Dr\"oge et al.}, 2010; {\it He et al.}, 2011; {\it Qin et al.}, 2013]. In this study, we employ the focused transport equation, which is a form of the Fokker--Planck equation, described as

\begin{equation}
	\partial_t f+\mu vb_i\partial_i f +V_i\partial_i f +\dot{p} \partial_p  f +\dot{\mu}\partial_\mu f -\partial_\mu \left(D_{\mu\mu}\partial_\mu f\right)=0,
\label{eq:fte}
\end{equation}
where $p$, $v$, and $\mu$ are the momentum intensity, speed, and pitch-angle cosine of a particle, respectively, $i$ denotes the three components of spatial coordinates, $b_i$ and $V_i$ are the unit vector along the magnetic field line and solar wind velocity, respectively, $D_{\mu\mu}$ denotes the pitch-angle scattering coefficient, $\partial$ with subscripts $t$, $i$, $p$, and $\mu$ denote partial differential operators with respect to time, spatial coordinates, momentum intensity, and pitch-angle, respectively\footnote{In a mathematically precise description, the partial differential operators with respect to spatial coordinates should be written as covariant differential operators, and covariant and contravariant vectors should be distinctively written in a curvilinear coordinate system. However, these differences in description are not distinguished in this article.}, and $\dot{p}$ and $\dot{\mu}$ denote the time derivatives of $p$ and $\mu$, respectively. The gyrotropic phase space density $f$ is a function of $i$, $p$, $\mu$, and $t$. On the left-hand side of equation (\ref{eq:fte}), the second to sixth terms describe particle streaming, convection induced by solar wind, adiabatic deceleration, the change in the pitch-angle due to adiabatic focusing and diverging solar wind, and pitch-angle scattering, respectively. The adiabatic deceleration and change in the pitch-angle in equation (\ref{eq:fte}) are, respectively, defined as [{\it Isenberg}, 1997]

\begin{equation}
	\dot{p}=p\left(\frac{1-3\mu^2}{2}b_ib_j\partial_iV_j-\frac{1-\mu^2}{2}\partial_iV_i\right),
\label{eq:momentum_change}
\end{equation}
and
\begin{equation}
	\dot{\mu}=\frac{1-\mu^2}{2}\left[-vb_i\partial_i\left(\ln B\right)-\mu\left(3b_ib_j\partial_iV_j-\partial_iV_i\right)\right],
\label{eq:pitch_change}
\end{equation}
where $B$ is the magnetic field intensity. The first term in the square brackets on the right-hand side of equation (\ref{eq:pitch_change}) describes the adiabatic focusing effect. In this set of equations, the particle position and momentum are measured in the co-rotating frame of the Sun and the solar wind frame, respectively.
\par

We employ a simple Parker spiral magnetic field with radial solar wind speed $V_r$ and angular speed of solar rotation $\Omega$ for an interplanetary magnetic field structure defined as 

\begin{equation}
	\mbox{\boldmath $B$}=B_e\left(\frac{r_e}{r}\right)^2\left(\mbox{\boldmath $e$}_r-\tan\Phi\mbox{\boldmath $e$}_\varphi\right),
\label{eq:parker}
\end{equation}
where $\Phi$ denotes the angle between the radial and magnetic field directions and is expressed as 
\begin{equation}
	\tan\Phi=\alpha r\sin\theta,
\end{equation}
where $\alpha$ is the ratio of $\Omega$ to $V_r$, $r$, $\theta$, and $\varphi$ are, respectively, the radial distance, polar angle, and phase angle in spherical coordinates, $\mbox{\boldmath $e$}_r$ and $\mbox{\boldmath $e$}_\varphi$ are unit vectors denoting the radial and phase angle directions, respectively, and $B_e$ is the radial component of the magnetic field at a specific distance $r_e$. \par

The pitch-angle scattering coefficient $D_{\mu\mu}$ is defined by a quasi-linear theory [{\it Jokipii}, 1966] with some modification to avoid zero scattering at $\mu=0$ [{\it Beeck and Wibberenz}, 1986; {\it Bieber et al.}, 1994] and is described as

\begin{equation}
	D_{\mu\mu}=D_0vp^{q-2}\left(|\mu|^{q-1}+h\right)\left(1-\mu^2\right),
\end{equation}
where $D_0$ is related to the intensity of the turbulent magnetic field and $q$ is the inertial-range spectral index of the turbulent magnetic field. Here, a Kolmogorov spectrum is assumed, i.e., $q=5/3$, and the modification constant $h$ is set to $0.2$ [{\it Qin et al.}, 2006]. The pitch-angle scattering coefficient is related to the parallel mean free path as\footnote{A small correction is required for the parallel mean free path defined by equation (\ref{eq:mfp_relation}) in the case of focused transport [{\it Danos et al.}, 2013 and references therein]. However, equation (\ref{eq:mfp_relation}) is used as the relation between the pitch-angle scattering coefficient and the parallel mean free path in SEP transport simulations in this article.}
\begin{equation}
	\lambda_\parallel=\frac{3v}{8}\int^{+1}_{-1}\frac{\left(1-\mu^2\right)^2}{D_{\mu\mu}}d\mu.
\label{eq:mfp_relation}
\end{equation}
The radial mean free path is defined as 

\begin{equation}
	\lambda_{rr}=\lambda_\parallel\cos^2\Phi,
\end{equation}
which is assumed to be constant across the entire interplanetary space [{\it Bieber et al.}, 1994]. \par

\subsection{Coordinate system}
\label{sec:coordinate}

\begin{figure}
	\includegraphics[width=\textwidth]{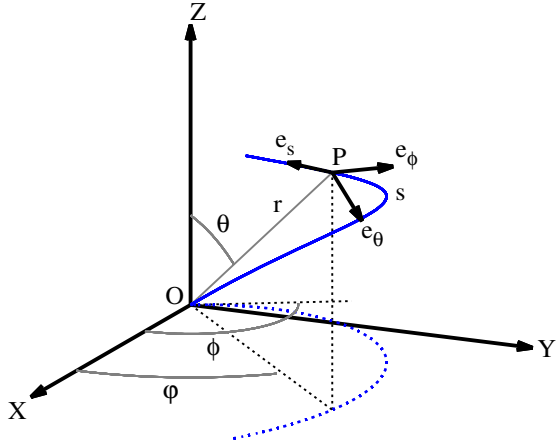}
	\caption{Non-orthogonal curvilinear coordinate system ($s$, $\theta$, $\phi$) for describing the focused transport equation. The solid and dotted blue lines are respectively a Parker magnetic field and its projection onto the X-Y plane, $s$ is the distance of $P$ measured from the origin $O$ along the magnetic field line, $\theta$ is the polar angle at $P$ measured from the $Z$ axis, $\phi$ is the phase angle of the magnetic field line at $O$ measured from the $X$ axis, which is different from the phase angle $\varphi$ at $P$, $r$ is the radial distance from $O$ to $P$, $\mbox{\boldmath $e$}_s$, $\mbox{\boldmath $e$}_\theta$, and $\mbox{\boldmath $e$}_\phi$ are the basis vectors at position $P$, and ($r$, $\theta$, $\varphi$) constitutes the spherical coordinate system.}
\label{fig:coordinate}
\end{figure}

As a convenient coordinate system for this problem, a magnetic field line coordinate system ($s$, $\theta$, $\phi$) is defined, as shown in Figure \ref{fig:coordinate}. In the figure, $s$ is the distance of position $P$ measured from the origin $O$ along a magnetic field line, $\theta$ is the polar angle at $P$ measured from the $Z$ axis, which is fixed at $\pi /2$ in this article, and $\phi$ is the phase angle of the magnetic field line at $O$, {\it rather than at} $P$, measured from the $X$ axis. The relationship between the magnetic field line coordinate system ($s$, $\theta$, $\phi$) and the spherical coordinate system ($r$, $\theta$, $\varphi$) is expressed as

\begin{align}
	s       &=\frac{r}{2}\left[\frac{1}{\cos\Phi}+\frac{1}{\sin\Phi}\ln\left(\tan\Phi+\frac{1}{\cos\Phi}\right)\right],
\label{eq:s_r}\\
	\theta&= \theta, 
\label{eq:theta}\\
	\phi   &= \alpha r+\varphi.
\label{eq:phi_varphi}
\end{align}
The magnetic field line coordinate system is identical to the spherical coordinate system when the solar rotation vanishes, i.e., when $\alpha=0$. \par

The solar wind velocity in the co-rotating frame of the Sun only has an $s$-component in the magnetic field line coordinate system, when is expressed as 

\begin{equation}
	V_s=\frac{V_r}{\cos\Phi}.
\end{equation}
\par

As the magnetic field line coordinate system is a non-orthogonal curvilinear coordinate system, off-diagonal elements appear in the metric tensor, and the line element $dl^2$ of the coordinate system is described as

\begin{equation}
	dl^2=ds^2+r^2d\theta^2+\left(r\sin\theta\right)^2d\phi^2-\frac{2}{\alpha}\tan^2\Phi\cos\Phi \ dsd\phi,
\label{eq:line_element}
\end{equation}
where the relationship between $r$, $s$, and $\theta$ is given by equation (\ref{eq:s_r}). Equation (\ref{eq:fte}) can be recast in the magnetic field line coordinate system as
\begin{equation}
	\partial_t f+\left(\mu v+V_s\right)\partial_s f +\dot{p} \partial_p f +\left(\dot{\mu}-\partial_\mu D_{\mu\mu}\right)\partial_\mu f -D_{\mu\mu}\partial_\mu\partial_\mu f=0,
\label{eq:fte2}
\end{equation}
where the rate of change of the momentum and $\dot{p}$ and rate of change of the pitch-angle $\dot{\mu}$ are, respectively, expressed as

\begin{equation}
	\dot{p}=p\frac{V_s\cos\Phi}{r}\left[\frac{1-3\mu^2}{2}\sin^2\Phi-\left(1-\mu^2\right)\right],
\label{p_dot}
\end{equation}
and
\begin{equation}
	\dot{\mu}=\frac{1-\mu^2}{2r}\cos^3\Phi\left[v\left(2+\tan^2\Phi\right)+\mu V_s\left(2-\tan^2\Phi\right)\right].
\label{mu_dot}
\end{equation}
Equation (\ref{eq:fte2}) can easily be solved because of the spatial one-dimensionality along a magnetic field line.
\par

\subsection{Method used to solve focused transport equation}
\label{sec:solve}
The focused transport equation described above is solved by solving a set of stochastic differential equations that is mathematically equivalent to the Fokker-Planck equation [{\it Kloeden and Platen}, 1999; {\it \O ksendal}, 1999; {\it Zhang}, 1999]. Stochastic differential equations are increasingly used to model phenomena such as SEP transport [e.g., {\it Zhang et al.}, 2009; {\it Dr\"oge et al.}, 2010; {\it Qin et al.}, 2013], cosmic-ray transport in the heliosphere [e.g., {\it Pei et al.}, 2010; {\it Strauss et al.}, 2013], cosmic-ray transport in the Galaxy and interstellar media [e.g., {\it Farahat et al.}, 2008; {\it Effenberger et al.}, 2012], and particle acceleration in shock waves [e.g., {\it Zuo et al.}, 2011]. \par

While the Fokker--Planck equation can generally be recast as a set of both time-forward and time-backward stochastic differential equations, we use a set of time-backward stochastic differential equations because a time-backward method is suitable for calculating the time evolution of particle intensity or the energy spectra of an SEP event observed near the Earth [{\it Kopp et al.}, 2012]. Therefore, equation (\ref{eq:fte2}) is recast as a set of time-backward stochastic differential equations as follows:

\begin{equation}
	ds=-\left(\mu v+V_s\right)d\tau,
\end{equation}

\begin{equation}
	dp=-\dot{p} d\tau,
\end{equation}

\begin{equation}
	d\mu =-\left(\dot{\mu}-\partial_\mu D_{\mu\mu}\right)d\tau+\sqrt{2D_{\mu\mu}}dw,
\label{eq:dmu}
\end{equation}
and

\begin{equation}
	d\tau=-dt,
\end{equation}
where $\tau$ is the time measured backward and $dw$ in equation (\ref{eq:dmu}) denotes a one-dimensional Wiener process, the probability density function of which is described by the Gaussian function

\begin{equation}
	p\left(dw;d\tau\right)=\frac{1}{\sqrt{2\pi d\tau}}\exp\left(-\frac{dw^2}{2d\tau}\right).
\label{eq:pdf}
\end{equation}
\par

To numerically solve this set of stochastic differential equations, the motion of pseudo particles can be tracked by considering numerous stochastic processes using a Monte Carlo method. In accordance with equation (\ref{eq:pdf}), random variables are generated for each step of numerical integration by a pseudo random number generator. Pseudo particles are numerically traced backward in time over a specified duration. When a particle comes in contact with an inner or outer boundary of the calculation domain, calculations for that particular particle are terminated. The inner and outer boundaries are set at heliocentric radial distances of $0.05$ and $80$ AU, respectively. \par

The expectation value of the generated stochastic processes weighted by an inner boundary condition is given by

\begin{equation}
	f\left(s,p,\mu,t\right)=\left<f_0\left(s_{t-\tau_0},p_{t-\tau_0},\mu_{t-\tau_0},t-\tau_0\right)\right>,
\end{equation}
where the brackets $\left<\cdots\right>$ denote the expectation value and $\tau_0$ is the backward time at which the pseudo particle comes in contact with the inner boundary. $f_0$ denotes the inner boundary condition, which expresses the SEP injection profile near the Sun. The expectation value $f(s,p,\mu,t)$ describes the phase space density at time $t$ of the focused transport equation, which is the final solution. \par

The SEP injection profile at the inner boundary is given by

\begin{equation}
	f_0\left(s,p,\mu,t\right)=\frac{\gamma-1}{2p_{min}}\left(\frac{p}{p_{min}}\right)^{-\gamma} \sqrt{\frac{m^3}{2\pi\sigma^2t^3}}\exp\left[-\frac{m\left(t-m\right)^2}{2\sigma^2t}\right],
\label{eq:injection}
\end{equation}	
and it is normalized to unity over the integral with ranges $p_{min}<p<\infty$, $0<\mu<1$, and $0<t<\infty$. At the time of injection, the pitch-angle is uniformly distributed between $0$ and $\pi/2$, and the momentum spectrum is a power law with index $-\gamma$. The time profile is expressed as an inverse Gaussian distribution with mean $m$ and standard deviation $\sigma$. Although most of the studies on interplanetary SEP transport [e.g., {\it Dr\"oge}, 2000; {\it Qin et al.}, 2006; {\it Zhang et al.}, 2009] use the so-called Reid-Axford profile [{\it Reid}, 1964] as the injection time profile, we employ the inverse Gaussian distribution as the injection time profile for the following reason. The inverse Gaussian distribution is related to the diffusion process as $dX=\lambda dt+\delta dw$, where $X$, $t$, and $dw$ denote the state variable, time, and Wiener process, respectively, $\lambda$ is the average rate of change of the state variable, and $\delta^2$ denotes the diffusion coefficient. The first and second terms on the right-hand side of the equation are called the drift and diffusion terms, respectively. The time $t$ at which $X$ first attains state $X_f$, starting from $X=X_0$ at $t=0$, has an inverse Gaussian distribution. In physical terms, if energetic particles are injected impulsively at a specific position in the lower corona and move outward at a constant drift rate accompanied by diffusion, the particles that escape at a specific distance in the upper corona have an inverse Gaussian distribution in time. Therefore, we adopt the inverse Gaussian distribution as the particle injection time profile. \par

Because the pseudo particles integrated backward in time from the Earth to the Sun are pulled toward the outer boundary by pitch-angle focusing, most pseudo particles cannot return to the Sun. To allow particles to return to the Sun, a method proposed by Qin {\it et al.} [2006] is used, in which the focused transport equation is rewritten to suppress pitch-angle focusing. The expectation value is then calculated as the weighted average of the stochastic processes in accordance with the rewritten equation. The weights are calculated while the stochastic processes are generated. Mathematically, the solutions obtained using this method are identical to those of the original transport equation. \par

\section{Test calculations}
\label{sec:results}
By solving the focused transport equation described in the previous section using the injection profile equation, we can simulate the SEP event intensity and anisotropy profile. The anisotropy $A(t)$ parallel to the magnetic field is related to the pitch-angle distribution as 

\begin{equation}
	A(t)=\frac{3\int_{-1}^{+1} \mu f(\mu,t) d\mu}{\int_{-1}^{+1} f(\mu,t) d\mu}.
\end{equation}
\par

\begin{figure}
	\includegraphics[width=\textwidth]{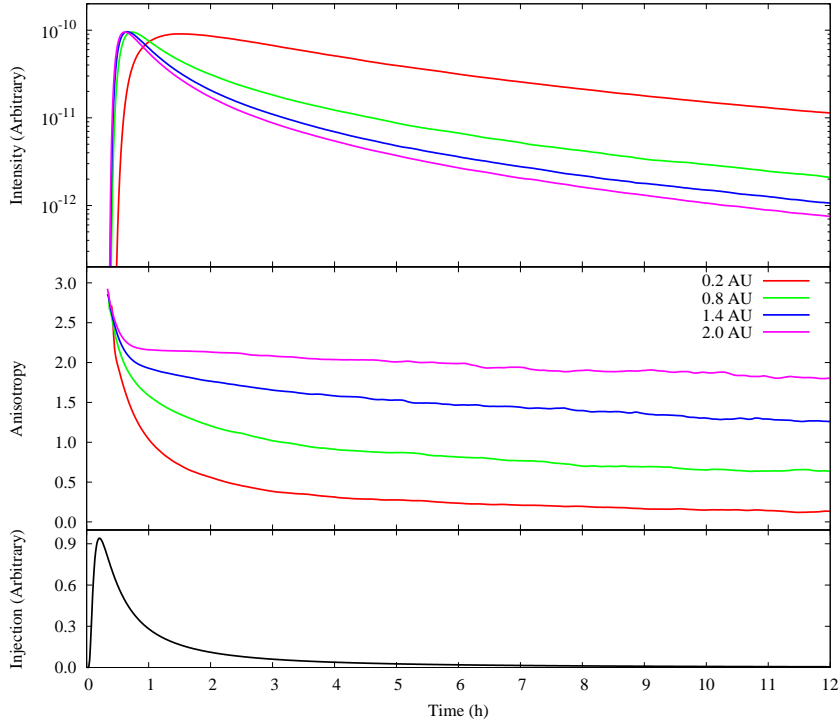}
	\caption{Simulated SEP intensity (top) and anisotropy (middle) near the Earth for radial mean free paths of 0.2 (red), 0.8 (green), 1.4 (blue), and 2.0 AU (pink). The bottom panel shows the particle injection profile expressed as an inverse Gaussian distribution with $m=2.5$ and $\sigma=5.0$ h.}
\label{fig:intensity_142}
\end{figure}

Figure \ref{fig:intensity_142} shows the simulated SEP intensity (top panel) and anisotropy (middle panel) for 142 MeV protons at distance of 1.0 AU from the Sun for radial mean free paths of 0.2, 0.8, 1.4, and 2.0 AU at 142 MeV, indicated as red, green, blue, and pink lines, respectively. The bottom panel shows the time profile of particle injection near the Sun with $m=2.5$, $\sigma =5.0$ hours, and $\gamma=5$, which are the same for all mean free paths. We can see from the figure that the SEP intensity and anisotropy strongly depend on the radial mean free path, namely the pitch-angle scattering. As shown in the top panel, the intensity decreases almost exponentially for the smallest mean free path (0.2 AU) and deviates further from exponential decay as the mean free path increases. As shown in the middle panel, the longer the mean free path, the slower the anisotropy decreases. This phenomenon demonstrates that more particles arrive from the solar direction as the mean free path increases because hardly any energetic particles are scattered back from behind the Earth. \par

\begin{figure}
	\includegraphics[width=\textwidth]{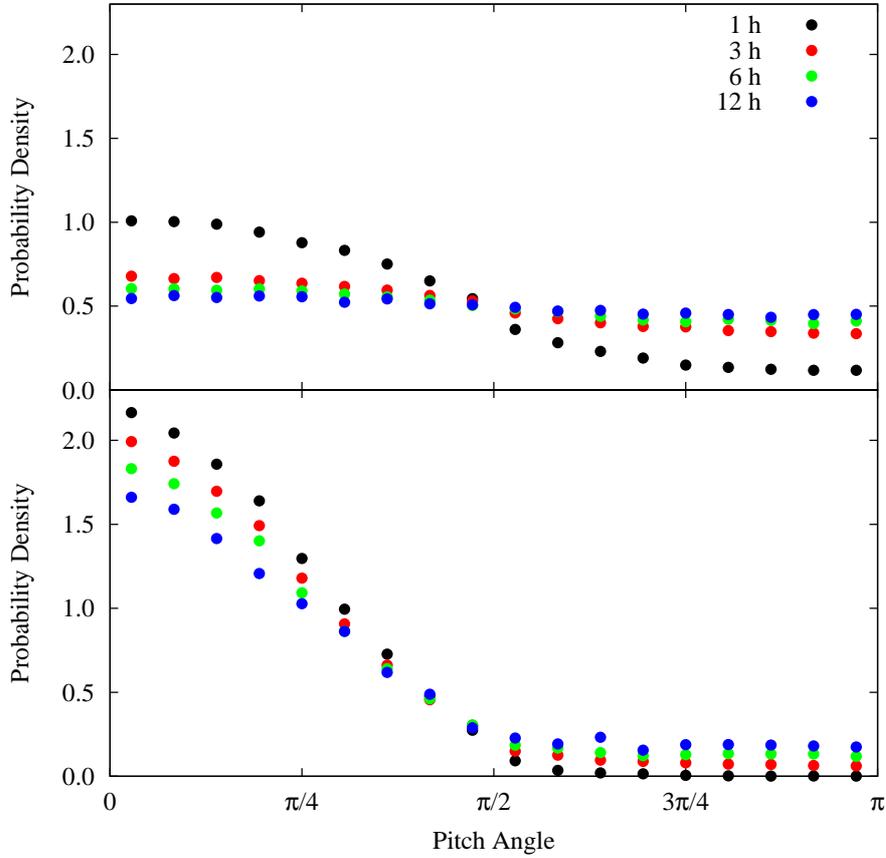}
	\caption{Pitch-angle distribution of near-Earth SEPs for radial mean free paths of 0.2 (top) and 1.4 AU (bottom). The distributions are shown at 1 (black), 3 (red), 6 (green), and 12 (blue) hours after particle injection near the Sun.}
\label{fig:pitch_142}
\end{figure}

Figure \ref{fig:pitch_142} shows the temporal evolutions of the pitch-angle distribution for mean free paths of 0.2 AU (top panel) and 1.4 AU (bottom panel), which are shown as red and blue lines in Figure \ref{fig:intensity_142}, respectively. The black, red, green, and blue points in the figure show the pitch-angle distributions at 1, 3, 6, and 12 h, respectively, following particle injection near the Sun. The pitch-angle is measured from the direction of the magnetic field line away from the Sun. In both cases, the pitch-angle gradually becomes more isotropic from its initial anisotropic distribution. For the 1.4 AU mean free path, the pitch-angle remains highly anisotropic even after 12 h, whereas that for the 0.2 AU mean free path is almost completely isotropic. This phenomenon is attributed to the fact that hardly any particles are scattered back from behind the Earth owing to the long mean free path or weak scattering in solar wind. \par

\begin{figure}
	\includegraphics[width=\textwidth]{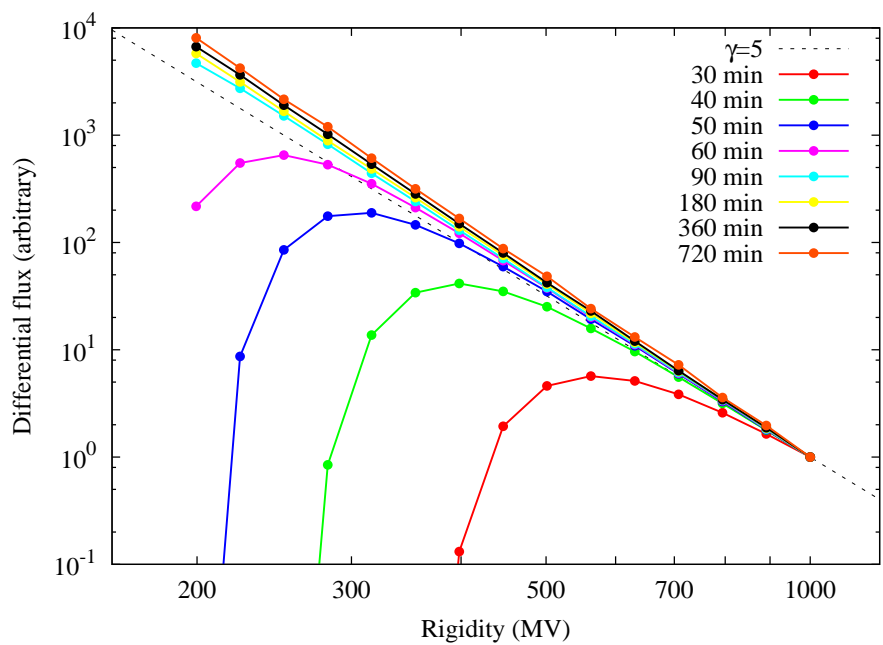}
	\caption{Calculated rigidity spectra near the Earth. The dashed line shows the injection spectrum at the Sun. Colored lines show the temporal evolutions of spectra at the Earth. The spectra are normalized using a value at 1,000 MV.}
\label{fig:spectrum}
\end{figure}

Figure \ref{fig:spectrum} shows calculated rigidity spectra at the Earth for a mean free path of 0.8 AU at 1 GV. The eight rigidity spectra shown in the figure are for times of 30, 40, 50, 60, 90, 180, 360, and 720 min after the start of particle injection near the Sun. The dashed line shows the injection spectrum at the Sun with a power-law index of -5. The injection parameters are the same as those for Figure \ref{fig:intensity_142}. It is evident that the rigidity spectra change with time. This complex behavior of the spectral shape is ascribed to the time-of-flight effect, which means that the particle fluxes in the low-rigidity range are low in the early phase because low-rigidity particles arrive near the Earth later than high-rigidity particles do. Our simulation can calculate the complex behavior of SEP momentum spectra, which is essential to calculate neutron monitor count rates during a GLE, for example. \par

\section{Comparison with observations}
\label{sec:comparison}
In this section, we compare observational data with the results of our numerical simulation. The data we used are differential intensities obtained at 1-min intervals with the GOES P6 and P7 channels, whose energies are 165 MeV and 433 MeV (580 MV and 1,000 MV, respectively in rigidity for protons). Because the GOES satellites are located inside the terrestrial magnetosphere, the observed SEP intensities may have been affected by the magnetosphere. However, this effect will be small at the high particle energies of the P6 and P7 channel data. To reduce the effect of the longitudinal difference between the two GOES satellites on the SEP intensities, the data collected from GOES 13 and 15 are averaged. Noise is eliminated by calculating the running average of data points. \par

The method of comparison is quite simple; the results of the numerical simulations are fitted to the observed data using a selected radial mean free path. Data are fitted using a nonlinear least-squares method and the square error is minimized using the simplex method [{\it Nelder and Mead}, 1965]. The mean $m$ and standard deviation $\sigma$ of the inverse Gaussian injection profile are adjusted to match the simulated intensity of the observed SEP event. Because the results are insensitive to the momentum spectrum index $\gamma$, it is fixed at 5. \par

\subsection{Events used for comparison}
\label{sec:event}
For reproducing extreme events, we selected events with intensities for SEPs having energy greater than 100 MeV observed by the GOES satellite reaching 1 PFU. Nine such events have been observed since 2012. By checking SDO/AIA $94-\mathrm{\AA}$ images, it was found that four of the nine events occurred in the eastern hemisphere or near the central meridian region of the Sun (longitude less than W30), and we discarded these four events. One of the remaining events had missing data for both GOES 13 and 15. Therefore, four events remained: those on 27th January 2012, 13th March 2012, 17th May 2012, and 6th January 2014. We compared the four events with the results of numerical simulations. \par

\subsection{Event on 27th January 2012}
\label{sec:27january2012}
\begin{figure}
	\includegraphics[width=\textwidth]{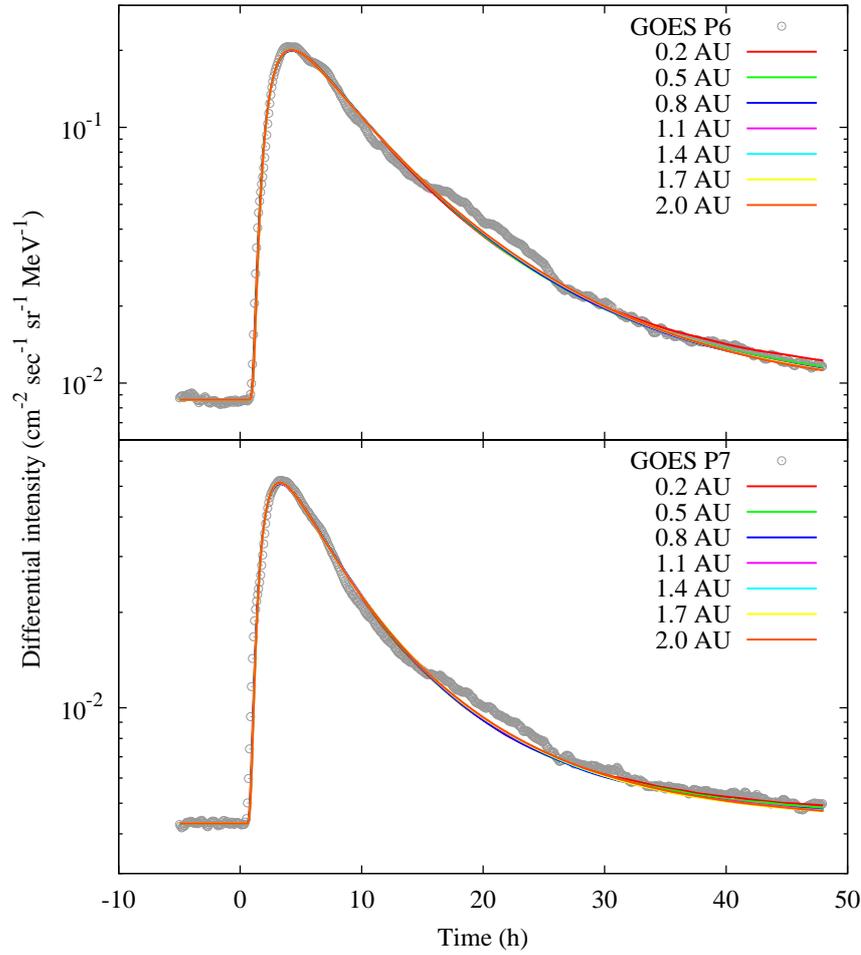}
	\caption{Comparison between calculated SEP intensity and the GOES P6 (upper panel) and P7 (lower panel) channel data for the event on 27th January 2012. The gray points are the observation data, which are plotted in the figure by thinning them out to increase their visibility. The colored lines show the SEP intensity profiles calculated using our method for various mean free paths.}
\label{fig:comparison201201}
\end{figure}

Figure \ref{fig:comparison201201} shows a comparison of the SEP event that occurred on 27th January 2012 with the results of our numerical simulation. The upper and lower panels represent the P6 and P7 channel data, respectively. The gray points indicate the observation data, which are plotted in the figure by thinning them out to increase their visibility. Unfortunately, as the GOES satellites do not observe the anisotropy of SEPs, we do not know the scattering mean free path in interplanetary space, and the comparisons should be performed over a range of possible radial mean free paths. Therefore, we compared the simulation results for the SEP intensities calculated using the proposed method (colored lines) for mean free paths ranging from 0.2 to 2.0 AU (increment: 0.3 AU) with the GOES data. The figure shows excellent agreement between the observational data for both the P6 and P7 channels and our simulation for all mean free paths. \par

\subsection{Event on 13th March 2012}
\label{sec:13march2012}
\begin{figure}
	\includegraphics[width=\textwidth]{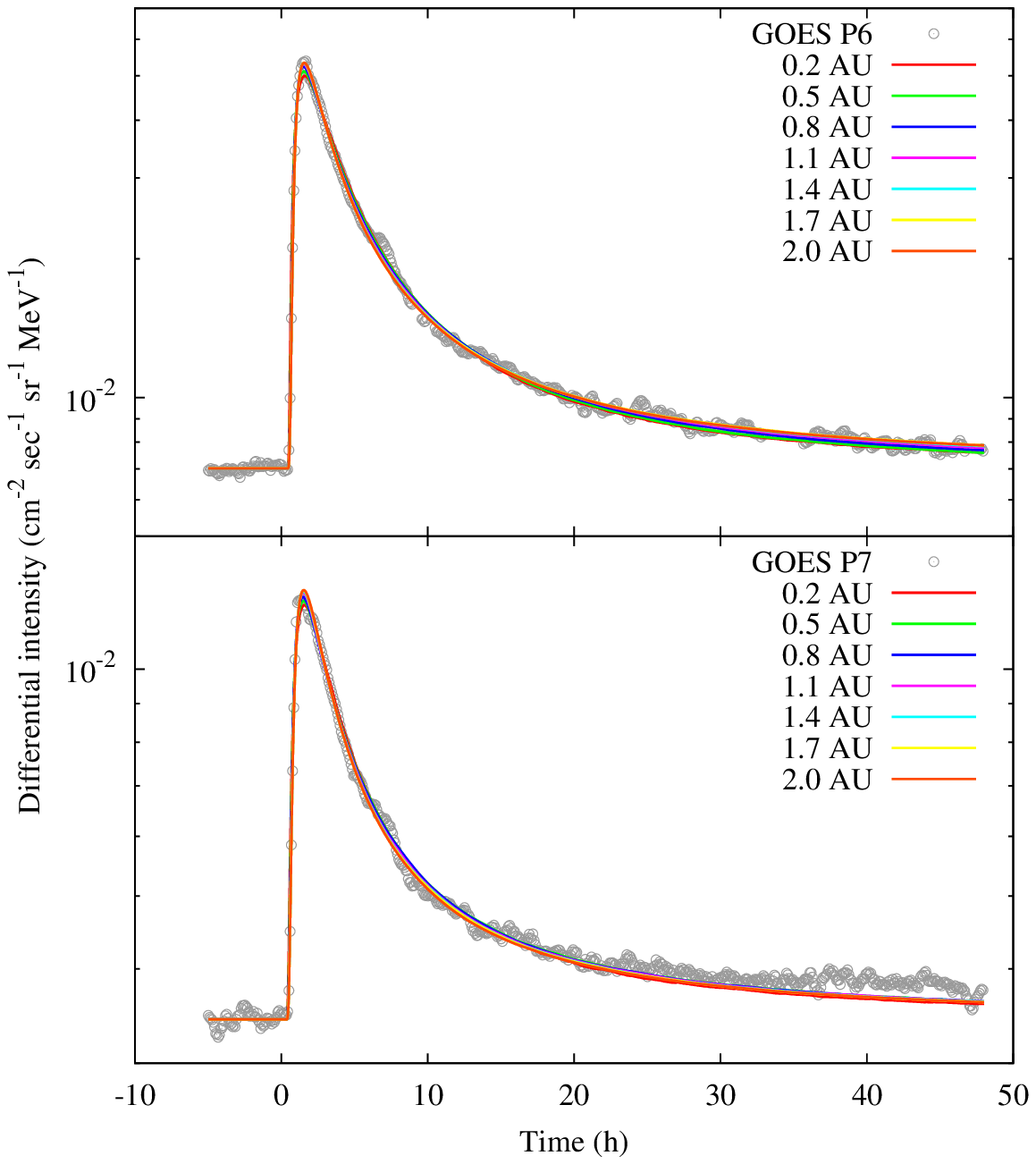}
	\caption{Same as Figure \ref{fig:comparison201201}, but for the SEP event occurring on 13th March 2012.}
\label{fig:comparison201203}
\end{figure}

 Figure \ref{fig:comparison201203} shows a comparison between the event observed with the GOES P6 and P7 channels on 13th March 2012 and the calculated intensity. The gray points and colored lines have the same meanings as those in Figure \ref{fig:comparison201201}. The upper panel shows that the P6 channel data are almost perfectly reproduced by our numerical simulation results. In the lower panel, the P7 channel data are again well reproduced, although the late phase data of the event deviate slightly from the simulation results. \par

\subsection{Event on 17th May 2012}
\label{sec:17may2012}
\begin{figure}
	\includegraphics[width=\textwidth]{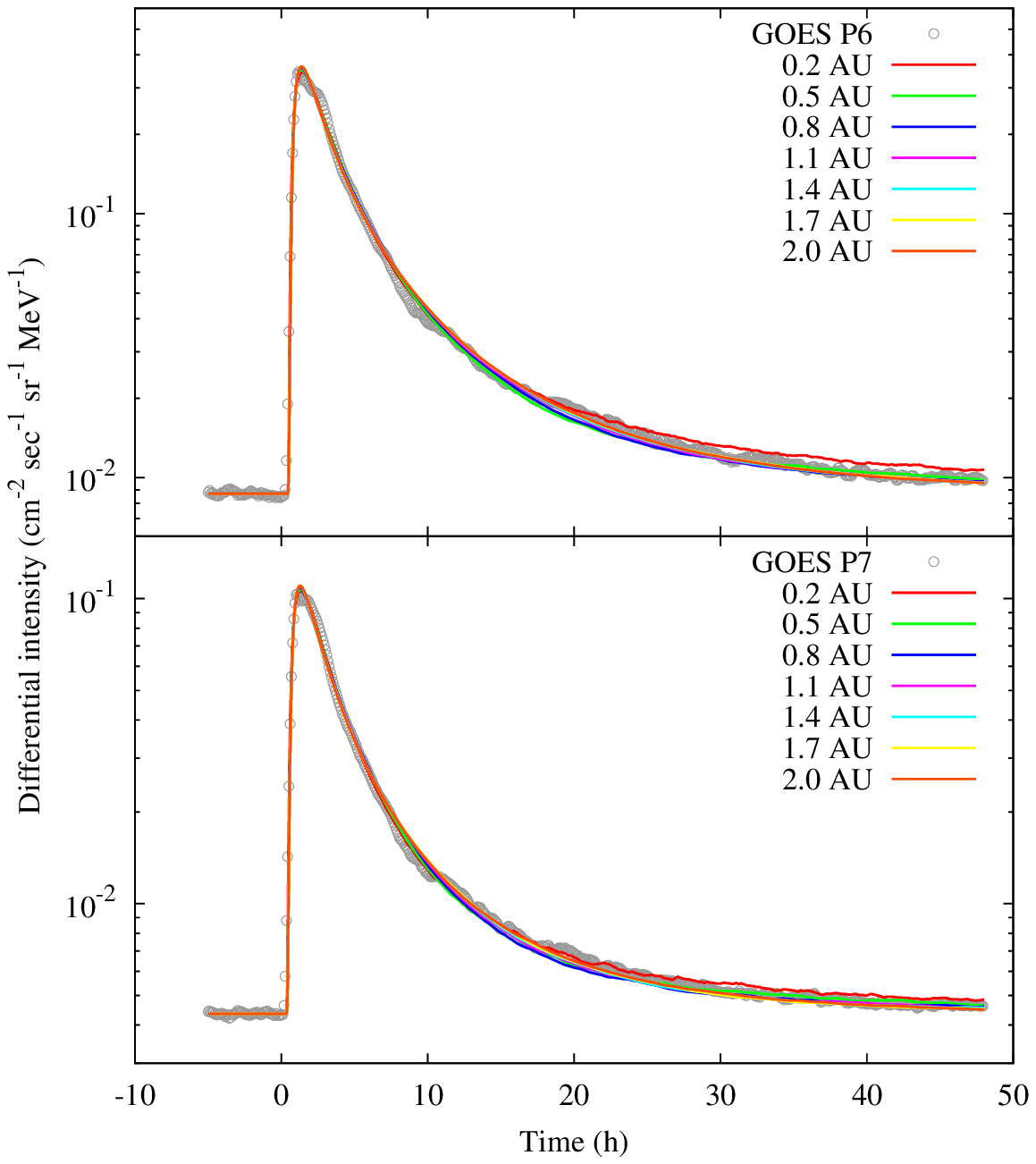}
	\caption{Same as Figure \ref{fig:comparison201201}, but for the SEP event occurring on 17th May 2012.}
\label{fig:comparison201205}
\end{figure}

The SEP event that occurred on 17th May 2012 was the first ground level enhancement (GLE) within the 24th solar cycle, which was an extreme space weather event. Figure \ref{fig:comparison201205} shows a comparison of the observed data with the calculated intensity. The gray points and colored lines have the same meanings as those in Figure \ref{fig:comparison201201}. Again, our simulation results reproduce the observed SEP event with high accuracy for all mean free paths and for both P6 and P7 data, even though the decay rate of this event is considerably faster than that of the event on 27th January 2012. \par

\subsection{Event on 6th January 2014}
\label{sec:6january2014}
\begin{figure}
	\includegraphics[width=\textwidth]{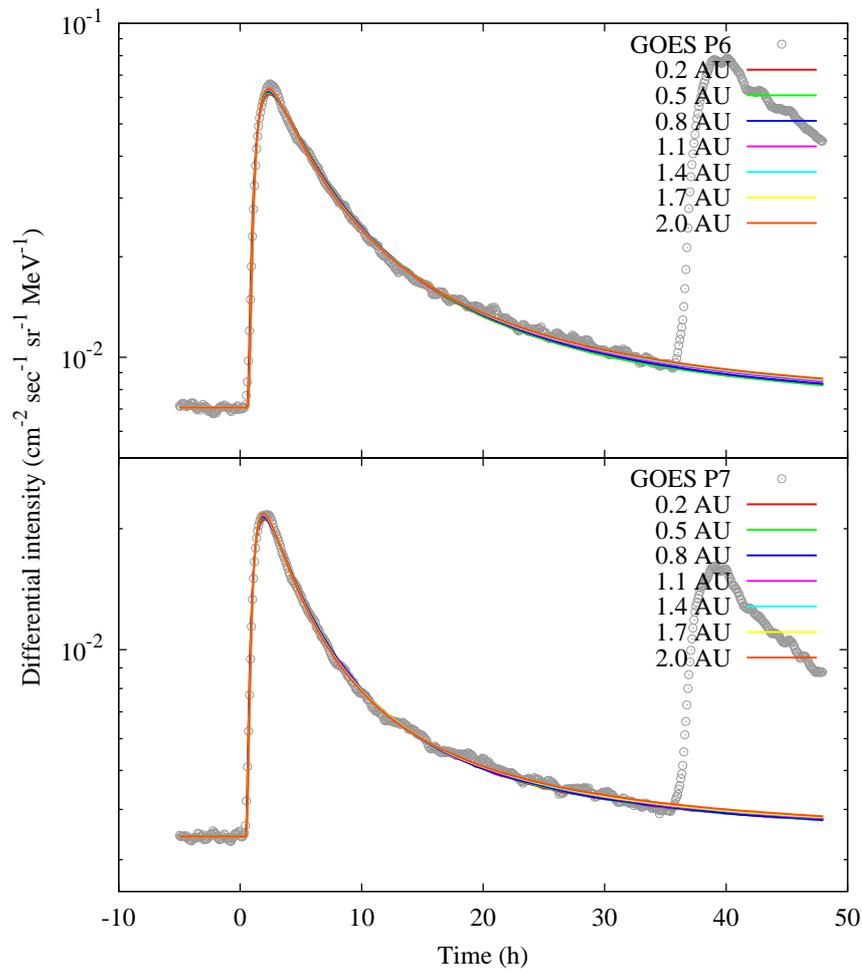}
	\caption{Same as Figure \ref{fig:comparison201201}, but for the SEP event occurring on 6th January 2014.}
\label{fig:comparison201401}
\end{figure}

The SEP event that occurred on 6th January 2014 was an interesting event. No increase in X-ray flux was detected by GOES at the time of the event. However, a flare was clearly observed in the eastern hemisphere from the view of STEREO-A, suggesting that the SEP event was related to a flare that occurred on the far side of the Sun. Figure \ref{fig:comparison201401} shows the fitting result for this event. Excellent agreement between the observation data and our calculations can be observed in both the upper (P6 channel) and lower (P7 channel) panels. As another SEP event occurred toward the end of this event, the data during this period were not used in the fitting procedure. \par

The comparisons shown in this section indicate that the inverse Gaussian time profile can be reasonable for the SEP injection profile near the Sun. \par

\section{Discussion}
\label{sec:discussion}
We used the inverse Gaussian distribution as the SEP injection profile in the solar corona in this study. As already mentioned, however, the Reid-Axford profile is often used for the injection profile. The time dependence of the exponential functions included in both the inverse Gaussian and Reid-Axford profiles is the same, and the only difference between them is the time dependence outside the exponential function. While the Reid-Axford profile has $t^{-1}$ time dependence, the inverse Gaussian profile has $t^{-1.5}$ time dependence. This difference is ascribed to the different physical mechanisms considered for the two profiles. While the Reid-Axford profile is derived on the basis of lateral diffusion in the solar corona, the inverse Gaussian is based on vertical drift accompanied by diffusion in the solar corona. In reality, accelerated particles can drift and diffuse anisotropically in three dimensions, and the degrees of drifting and diffusion along each direction depend on the coronal magnetic field configuration, therefore, the injection model that is more suitable depends on the event. \par

\begin{figure}
	\includegraphics[width=\textwidth]{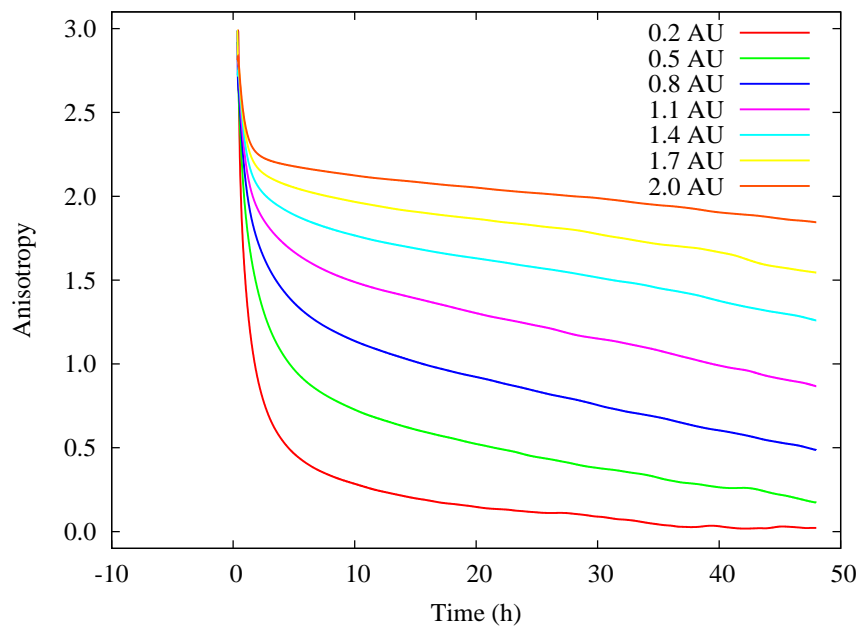}
	\caption{Calculated anisotropy profiles for the event occurring on 27th January 2012. From bottom to top, the lines are plotted for mean free paths ranging from 0.2 to 2.0 AU in 0.3 AU increments.}
\label{fig:anisotropy201201}
\end{figure}

As mentioned in the previous section, an identical SEP injection profile yields different intensity profiles depending on the scattering mean free path in interplanetary space. The mean free path is a crucial parameter for determining the SEP transport conditions. However, the comparison shown in the previous section predicted almost identical SEP intensity profiles over a wide range of mean free paths. This is because, in the comparison the injection profile was adjusted to fit the observational data for each mean free path. In contrast, the anisotropy profiles are widely scattered for similar calculated intensity profiles and different mean free paths. Figure \ref{fig:anisotropy201201} illustrates the calculated anisotropy profiles for the event on 27th January 2012. The lines from bottom to top represent the anisotropy profiles for radial mean free paths ranging from 0.2 to 2.0 AU with increments of 0.3 AU. The anisotropy varies considerably with the mean free path, although the calculated intensity profiles are similar. Although no observation of the anisotropy has yet been reported for SEPs with energy greater than several tens of MeV, if such observations become available, our model may be able to estimate the scattering mean free path in interplanetary space in near real-time, which will be valuable for the study of SEP transport and for further predicting the SEP intensity. Thus, anisotropy observations for SEPs, particularly with energy greater than several tens of MeV, will be valuable for SEP research and prediction.\par

\begin{figure}
	\includegraphics[width=\textwidth]{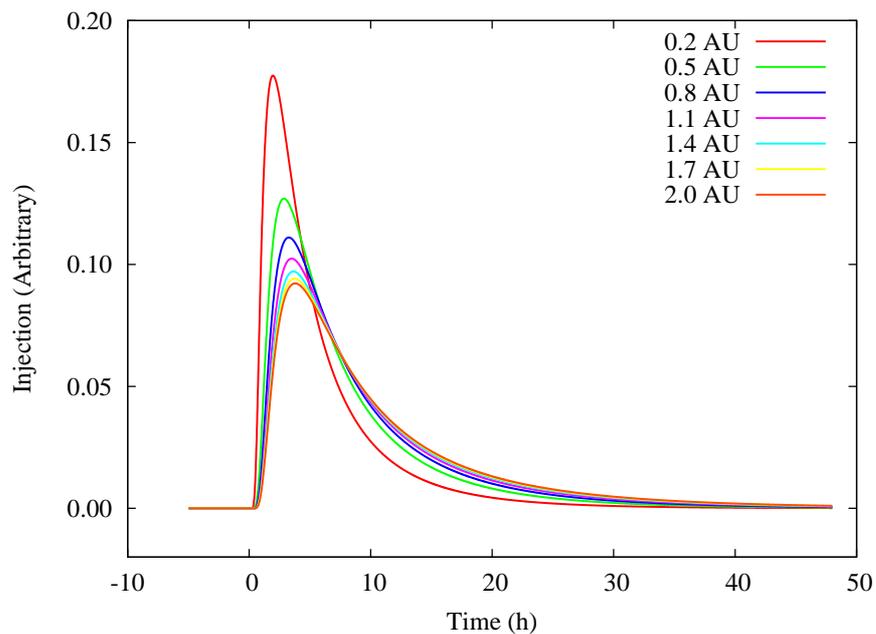}
	\caption{Derived injection profiles for the event occurring on 27th January 2012. The lines are plotted for mean free paths ranging from 0.2 to 2.0 AU in 0.3 AU increments.}
\label{fig:injection201201}
\end{figure}

Figure \ref{fig:injection201201} illustrates the derived injection profiles for the event on 27th January 2012. Colored lines represent the injection profiles for radial mean free paths ranging from 0.2 to 2.0 AU with increments of 0.3 AU. We can recognize from the figure that the longer the mean free path, the longer the injection time scale. A long mean free path such as 2.0 AU results in a quite long injection time. One possible interpretation of the long injection time is that the SEP source is far from the magnetic foot point of the Earth. In this case, the SEPs must diffuse across the magnetic field lines in the solar corona to arrive at the magnetic field lines connected to the Earth. It takes the SEPs a long time to escape from the upper corona to interplanetary space. \par

Because the simulation is spatially one-dimensional, perpendicular diffusion and gradient-curvature drift are not included, and particles only move along magnetic field lines. This means that particle acceleration sites must be connected to observation sites via magnetic field lines, which may limit the applicability of the method to well-connected events. All events analyzed in the previous section were accompanied by solar flares in the western hemisphere of the Sun, suggesting that these events were well-connected. On the other hand, according to {\it Gopalswamy et al.} [2013], the event on 27th January 2012 was not well-connected because of the large latitudinal separation ($\sim 40^\circ$) between the CME location and the ecliptic plane, although this event occurred in the western hemisphere of the Sun. This is why the initial phase of the event is slightly slow-rising. As already mentioned, the event on 6th January 2014 occurred at the far-western side of the Sun, implying that this event was also not well-connected. Therefore, our simulation sometimes works well by choosing a slow-rising injection time profile even though the SEP events are not well-connected, as long as the accompanying flare/CME occurs in the western hemisphere of the Sun. \par

Our model may not reproduce events with a plateau around the intensity peak, such as the event on 29th September 1989, which occurred behind the western limb. This is because these events often include interplanetary-shock-accelerated SEPs, while our model only deals with SEPs accelerated near the Sun (at a solar flare and/or in the solar corona). As an interplanetary shock will seldom accelerate SEPs up to several tens of MeV (although an extremely strong interplanetary shock can achieve this), most high-energy SEPs, such as those observed by the GOES P6 and P7 channels, are accelerated near the Sun, and the intensity time profiles observed by the GOES P6 and P7 channels do not show a plateau. Actually, while the GOES P1 to P5 channel data for the 29th September 1989 event showed plateau profiles, the P6 and P7 channels showed decay profiles rather than plateau profiles. As this study focused on high-energy SEPs, such as those observed by the GOES P6 and P7 channels, our model will work well even for the event on 29th September 1989. From 1997 to 2014, 29 events in the western hemisphere (longitude greater than W30) with intensities of SEPs with energy greater than 100 MeV reaching 1 PFU were observed by the GOES satellite. By visually checking the GOES P6 and P7 channel data for the 29 events, we concluded that 23 of the 29 events can be fitted using our model (almost $80 \%$ of the western hemisphere events). \par

\begin{figure}
	\includegraphics[width=\textwidth]{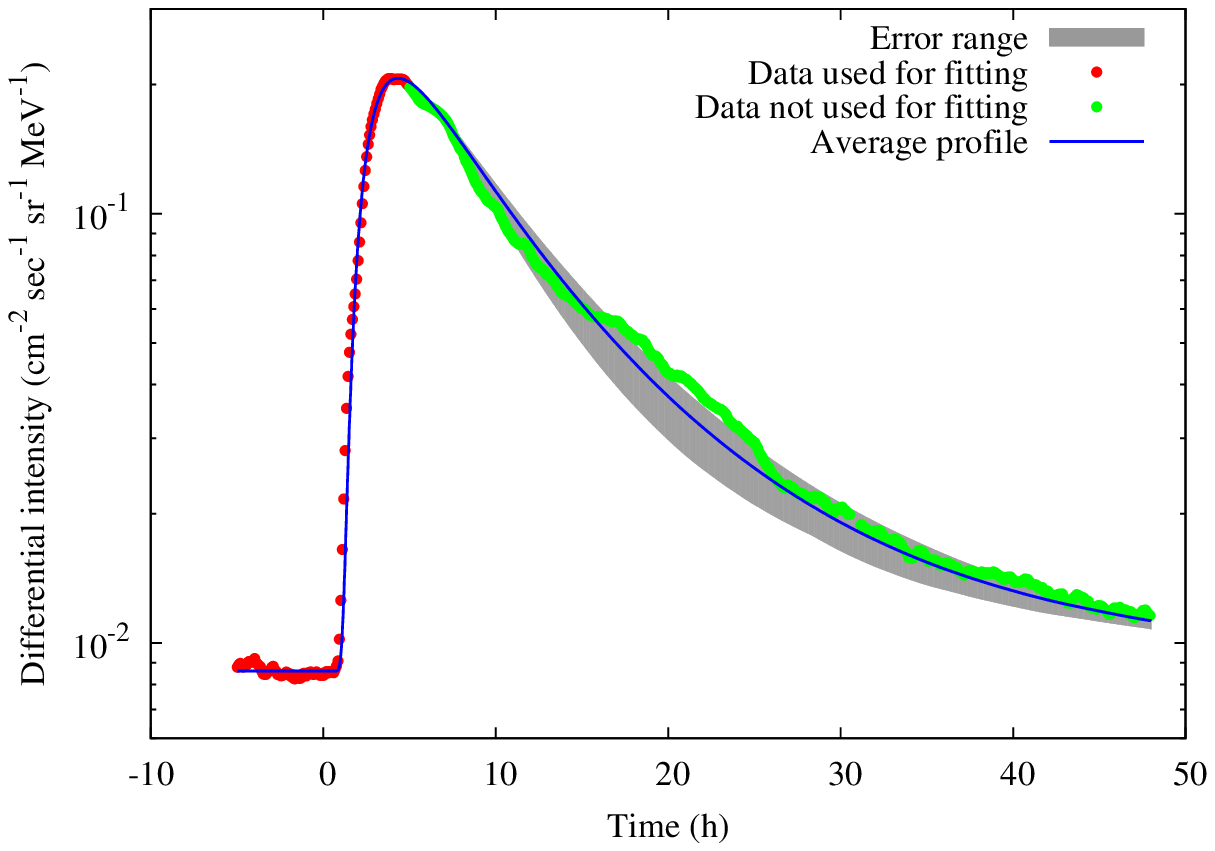}
	\caption{Results of GOES P6 channel data fitting using only the initial phase data of the 27th January 2012 event. The red and green points are the observation data used and those not used for the fitting, respectively. The blue line shows the SEP intensity profile calculated using our simulation, which was averaged over the results for mean free paths ranging from 0.2 to 2.0 AU with increments of 0.3 AU. The gray area shows the error range (see text for details).}
\label{fig:pred201201}
\end{figure}

As an application of the numerical simulation, we may apply the calculated results to predict the decay phase time profile of an SEP event. We fitted the results of the calculation to the observed initial phase data of the GOES P6 channel with radial mean free paths ranging from 0.2 to 2.0 AU with increments of 0.3 AU. The mean $m$ and standard deviation $\sigma$ of the inverse Gaussian injection profile were adjusted to match the simulated intensity of the observed initial phase of the SEP event on 27th January 2012. After the fitting procedure, the decay phase intensity of the SEP event was calculated using the adjusted parameters $m$ and $\sigma$. The results are shown in Figure \ref{fig:pred201201}. The red and green points indicate the observation data used and those not used for fitting, respectively. Because the mean free path influences the predicted profile, the plotted predicted profile (blue line) is the average of the profiles predicted at different mean free paths. The upper and lower envelopes of all predicted profiles can be regarded as the upper and lower limits of the prediction, respectively. Thus, the region between these limits (gray region) can be regarded as the error bars or error range of the prediction. Because most of the observed data not used for fitting (green points) are located inside the error range in Figure \ref{fig:pred201201}, our calculations reproduce the decay phase of this event with a small error range. This is one of the applications of our numerical simulation, suggesting the possibility of using the SEP transport simulation to predict SEP decay phase time profiles. \par

As mentioned in the previous section, our simulation can closely reproduce GOES P7 channel data. The energy of the P7 channel is just the energy that produces an extreme event such as a GLE. Therefore, our simulation is expected to be useful for reproducing and/or predicting GLEs and radiation doses for astronauts and aircrews. \par

\section{Summary}
\label{sec:summary}
We developed a numerical simulation code to calculate SEP intensity profiles as one component of a recently developed warning system for aviation exposure to SEPs. The code solves a spatially one-dimensional focused transport equation recast as a set of stochastic differential equations. An inverse Gaussian distribution was employed instead of the conventionally used profile for the injection time profile of SEPs near the Sun. The code provides the temporal profile of the differential intensity of an SEP event. To validate the code, the simulation results obtained using the code were fitted to observational data for four SEP events on 27th January 2012, 13th March 2012, 17th May 2012, and 6th January 2014 that were recorded by the GOES P6 and P7 channels, the square error of the fitting was minimized using the simplex method. The model predicted all the event profiles with satisfactory accuracy, indicating that the inverse Gaussian distribution can be used as the SEP injection profile instead of the well-known injection profile. By successfully reproducing SEP events, this code is expected to be applied to develop the WASAVIES warning system for aviation exposure to solar energetic particles. \par

\newpage

%
%

\begin{acknowledgements}
The authors would like to acknowledge the National Geophysical Data Center, National Oceanic and Atmospheric Administration, for the use of GOES energetic particle data. We would also like to thank the anonymous referees for their useful comments, which helped us to improve the manuscript.
\end{acknowledgements}



\end{document}